\documentclass[aps,amsmath,showpacs,amsfonts,10pt]{revtex4}
\usepackage{epsfig,graphicx}
\usepackage[english]{babel}
\usepackage{amsfonts}
\usepackage{amsmath}
\usepackage{latexsym}
\usepackage{graphics,bm}
\usepackage{natbib}
\usepackage{dcolumn}
\usepackage{bm}
\usepackage{rotating}

\begin{document}

\title{Quantum dynamics of two-optical modes and a single mechanical mode optomechanical system: selective energy exchange }

\author{ Aranya B Bhattacherjee $^{1}$ and Neha Aggarwal $^{1,2}$}

\address{$^{1}$ Department of Physics, A.R.S.D College, University of Delhi (South Campus), New Delhi-110021, India}
\address{$^{2}$Department of Physics and Astrophysics, University of Delhi, Delhi-110007, India}

\begin{abstract}
We study the quantum dynamics of an optomechanical setup comprising two optical modes and one mechanical mode. We show that the same system can
undergo a Dicke-Hepp-Lieb superradiant type phase transition. We found that the coupling between the momentum quadratures of the two optical fields give rise to a new critical point. We show that selective energy exchange between any two modes is possible by coherent control of the coupling parameters. In addition we also demonstrate the occurrence of Normal Mode Splitting (NMS) in the mechanical displacement spectrum.
\end{abstract}


\maketitle

\section{Introduction}
Electromagnetic coupling between optical cavities and nano-mechanical resonators has led to new quantum mechanical behaviour of macroscopic quantum systems \citep{marquardt}. Substantial experimental progress has been made to realize such novel quantum systems at the single photon level \citep{murch,brennecke,purdy,teufel,chan,verhagen}. At the same time several theoretical works have appeared in the literature that study a single optical mode coupled to a single mechanical mode, in the strong coupling regime. Quantum effects are found in these hybrid systems if the coupling between the light field and the mechanical resonator becomes much larger than the cavity decay rate and the mechanical oscillation frequency. Quantum nonlinearities can also be introduced into the system by placing a Kerr medium inside the optical cavity \citep{tarun}.

It has been demonstrated experimentally that by coupling two optical modes to a mechanical oscillator, the quantum nonlinearity can be enhanced significantly \citep{thompson,grudinin,safavi,ludwig}. This enhanced nonlinearity has potential application in quantum nondemolition (QND) measurement of phonon number \citep{grangier}. Besides QND measurements, optomechanical systems have potential applications in quantum information processing \citep{stannigel}. The kerr type nonlinearity can form the basis of all optical switch with application to engineer a quantum phase gate for photonic or phononic qubits. The mechanical mode can also serve as a quantum memory \citep{chang}. The interaction between the optical and mechanical degrees of freedom gives rise to a quantum interface between solid-state, optical and atomic qubits \citep{stannigel2}. Optical modes in quantum information processing units can transfer information over long distances and on the other hand acoustic excitations (phonons) can store information for an extended period. A hybrid architecture composed of optical modes and mechanical modes can thus be fruitfully utilized to design a quantum communication and quantum information processing unit which can  store and transfer information coherently. Theoretical studies related to two mode system have been performed much in detail \citep{jayich,miao,gangat,lambert} .

Given these new developments in this field, in this work, we show that the dynamics of the linearized Hamiltonian of the two optical modes coupled to one mechanical mode is equivalent to that of the Dicke model \citep{emary}. We analytically calculate expressions for the normal modes of the system and show that the system can undergo a Dicke-Hepp-Lieb type superradiant phase transition by varying the various coupling constants. We demonstrate the possibility of selective energy exchange between any two modes. We also study the squeezing variances of the three modes near the quantum critical point. We demonstrate the normal mode splitting (NMS) in the displacement spectrum of the mechanical oscillator.

\section{Dicke-Hepp-Lieb Superradiant phase transition model}

The optomechanical setup that we consider here (Fig.1) is composed of an optical cavity with two optical modes (denoted by operators $a_{1}$ and $a_{2}$) and one mechanical mode (denoted by $b$). This system is described by the Hamiltonian \citep{safavi,ludwig}

\begin{figure}[h]
\hspace{-0.0cm}
\includegraphics [scale=0.80]{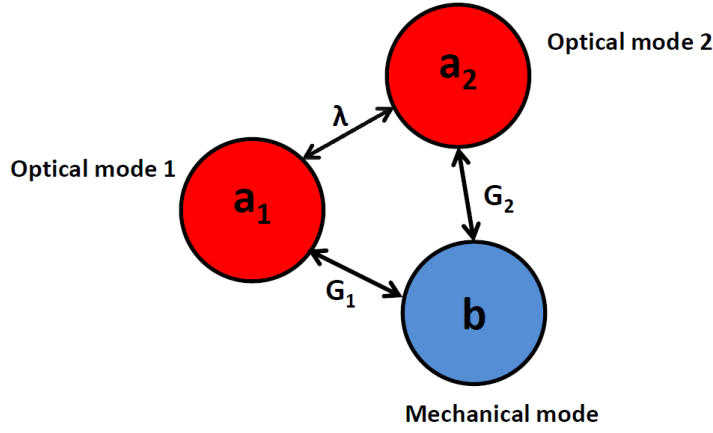}
\caption{Schematic representation of the hybrid optomechanical system consisting of two optical modes denoted by $a_{1}$, $a_{2}$ and one mechanical mode $b$. Also shown are the coupling rates between the various modes.  }
\label{f1g.1}
\end{figure}

\begin{equation}\label{ham1}
H=H_{0}+H_{int},
\end{equation}

where

\begin{equation}
H_{0}=\hbar \omega_{1} a^{\dagger}_{1} a_{1}+\hbar \omega_{2} a^{\dagger}_{2} a_{2}+\hbar \omega_{m} b^{\dagger}b,
\end{equation}

\begin{equation}
H_{int}=- \left [\hbar g_{1}a^{\dagger}_{1} a_{1} +\hbar g_{2} a^{\dagger}_{2} a_{2}-\hbar G (a_{1}a^{\dagger}_{2}+ a_{2} a^{\dagger}_{1}) \right ](b+b^{\dagger}).
\end{equation}

Here $H_{0}$ is the bare energies of the two optical modes with frequencies $\omega_{1,2}$ and that of the mechanical mode with frequency $\omega_{m}$. The term $H_{int}$ denotes the interaction between each of the two optical modes and the mechanical oscillator with coupling rates $g_{1,2}$. The coupling between the two optical modes via the mechanical oscillator is $G$. This kind of Hamiltonian is found typically in a setup with a membrane in the middle of the cavity \citep{thompson}. The interaction Hamiltonian results from the space dependence of the cavity mode frequencies.

We are interested in studying the dynamics of fluctuations of the system. To this end, we rewrite the operators $a_{1,2}$ and $b$ around their mean values as, $a_{1,2}\rightarrow \alpha_{1,2}+a_{1,2}$ and $b \rightarrow \beta +b$. Here $\alpha_{1,2}$ and $\beta$ are the mean field values of the optical modes and the mechanical mode respectively whose values are given in the Appendix I. As a result, we get an effective Hamiltonian for the system after retaining terms which are bilinear.

\begin{equation}\label{ham2}
H=\hbar \Omega_{1} a^{\dagger}_{1} a_{1}+\hbar \Omega_{2} a^{\dagger}_{2} a_{2}+\hbar \omega_{m} b^{\dagger}b - \left [\hbar G_{1}(a^{\dagger}_{1}+ a_{1}) +\hbar G_{2} (a^{\dagger}_{2}+ a_{2}) \right ](b+b^{\dagger})+\hbar \lambda (a_{1}a^{\dagger}_{2}+ a_{2} a^{\dagger}_{1}),
\end{equation}

where $\Omega_{1,2}=\omega_{1,2}-2 \beta g_{1,2}$, $G_{1,2}=g_{1,2} \alpha_{1,2}-G \alpha_{2,1}$ and $\lambda=2 G \beta$.

The Hamiltonian Eqn.(4) is bilinear in the fluctuation operators and can be easily diagonalized. This is accomplished by the following position and momentum operators for the three bosonic modes.

\begin{equation}
x=\frac{1}{\sqrt{2 \Omega_{1}}}(a^{\dagger}_{1}+a_{1}), \\\ p_{x}= i \sqrt{\frac{\Omega_{1}}{2}}(a^{\dagger}_{1}-a_{1})
\end{equation}

\begin{equation}
y=\frac{1}{\sqrt{2 \Omega_{2}}}(a^{\dagger}_{2}+a_{2}),\\\ p_{y}= i \sqrt{\frac{\Omega_{2}}{2}}(a^{\dagger}_{2}-a_{2})
\end{equation}

\begin{equation}
z=\frac{1}{\sqrt{2 \omega_{m}}}(b^{\dagger}+b), \\\ p_{z}= i \sqrt{\frac{\omega_{m}}{2}}(b^{\dagger}-b)
\end{equation}

The Hamiltonian (after ignoring constant terms) in terms of these operators is written as

\begin{equation}
H = \frac{1}{2} \left[\Omega_{1}^{2} x^{2}+p_{x}^{2}+\Omega_{2}^{2} y^{2} +p_{y}^{2}+ \omega_{m}^{2} z^{2} + p_{z}^{2}-4 G_{1} \sqrt{\Omega_{1} \omega_{m}}  x  z-  4 G_{2} \sqrt{\Omega_{2} \omega_{m}}  y z
+2 \lambda \left(\sqrt{\Omega_{1} \Omega_{2}}  x y + \frac{p_{x} p_{y}}{\sqrt{\Omega_{1} \Omega_{2}}}\right) \right].
\end{equation}

We now rotate the coordinate system in the following way:

\begin{equation}
x=q_{1} (\cos{\gamma_{1}}+\cos{\gamma_{2}})+q_{2} \sin{\gamma_{1}}+q_{3} \sin{\gamma_{2}},
\end{equation}

\begin{equation}
y=- q_{1} \sin{\gamma_{1}} +q_{2} (\cos{\gamma_{1}}+\cos{\gamma_{3}})+q_{3} \sin{\gamma_{3}},
\end{equation}

\begin{equation}
z=- q_{1} \sin{\gamma_{2}} - q_{2} \sin{\gamma_{3}}+q_{3} (\cos{\gamma_{2}}+\cos{\gamma_{3}}),
\end{equation}

where the angles $\gamma_{1}$, $\gamma_{2}$ and $\gamma_{3}$ are given by

\begin{equation}
\tan{2 \gamma_{1}}= \frac{2 \lambda \sqrt{\Omega_{1} \Omega_{2}}}{\Omega_{2}^{2}-\Omega_{1}^{2}},\ \tan{2 \gamma_{2}}= \frac{4 G_{1} \sqrt{\Omega_{1} \omega_{m}}}{\Omega_{1}^{2}-\omega_{m}^{2}},\ \tan{2 \gamma_{3}}= \frac{4 G_{2} \sqrt{\Omega_{2} \omega_{m}}}{\Omega_{2}^{2}-\omega_{m}^{2}}.
\end{equation}

This rotation eliminates the $xy$ , $xz$ and $yz$ interaction terms in the Hamiltonian. In order to eliminate the $p_{x} p_{y}$ interaction term, we further make the following transformation,

\begin{equation}
p_{x}= p_{1} \cos{\beta}+ p_{2} \sin{\beta}, \ p_{y}=-p_{1} \sin{\beta}+p_{2} \cos{\beta},
\end{equation}

where the angle $\beta$ is given as $\cos{2 \beta} =0 $. The Hamiltonian then takes the form of three uncoupled oscillators.

\begin{equation}
H=\frac{1}{2} \left\{ \frac{\epsilon_{x}^{2} q_{1}^{2}}{2}+ \frac{\epsilon_{y}^{2} q_{2}^{2}}{2}+\frac{\epsilon_{z}^{2} q_{3}^{2}}{2} + \epsilon_{p_{1}}p_{1}^{2}+ \epsilon_{p_{2}}p_{2}^{2}+ p_{3}^{2} \right\},
\end{equation}

where

\begin{equation}\label{eqn.x}
\epsilon_{x}=\sqrt{ \left\{ \frac{1}{2} \left[ \left(2 \Omega_{1}^{2}+\Omega_{2}^{2}+\omega_{m}^{2} \right) + \left( \sqrt{(\Omega_{1}^{2}-\omega_{m}^{2})^{2}+ 16 G_{1}^{2} \Omega_{1} \omega_{m}}-\sqrt{(\Omega_{2}^{2}-\Omega_{1}^{2})^{2}+ 4 \lambda^{2} \Omega_{1} \Omega_{2}}    \right)      \right]            \right\}},
\end{equation}

\begin{equation}\label{eqn.y}
\epsilon_{y}=\sqrt{ \left\{ \frac{1}{2} \left[ \left( \Omega_{1}^{2}+2 \Omega_{2}^{2}+\omega_{m}^{2} \right) + \left( \sqrt{(\Omega_{2}^{2}-\Omega_{1}^{2})^{2}+ 4 \lambda^{2} \Omega_{1} \Omega_{2}}+\sqrt{(\Omega_{2}^{2}-\omega_{m}^{2})^{2}+ 16 G_{2}^{2} \omega_{m} \Omega_{2}}    \right)      \right]            \right\}},
\end{equation}

\begin{equation}\label{eqn.z}
\epsilon_{z}= \sqrt{ \left\{ \frac{1}{2} \left[ \left( \Omega_{1}^{2}+\Omega_{2}^{2}+ 2 \omega_{m}^{2} \right) - \left( \sqrt{(\Omega_{1}^{2}-\omega_{m}^{2})^{2}+ 16G_{1}^{2} \Omega_{1} \omega_{m}}+\sqrt{(\Omega_{2}^{2}-\omega_{m}^{2})^{2}+ 16 G_{2}^{2} \omega_{m} \Omega_{2}}    \right)      \right]            \right\}},
\end{equation}

\begin{equation}
\epsilon_{p_{1}}= 1-\frac{\lambda}{\sqrt{\Omega_{1} \Omega_{2}}}, \ \epsilon_{p_{2}}= 1+ \frac {\lambda}{\sqrt{\Omega_{1} \Omega_{2}}}.
\end{equation}

We now requantize $H$ by introducing the following three bosonic modes,

\begin{equation}
q_{1}=\sqrt{\frac{\sqrt{\epsilon_{p_{1}}}}{2 \epsilon_{x}}} \left[ c_{1}^{\dagger} + c_{1} \right], \ p_{1}= i \sqrt{\frac{\epsilon_{x}}{2 \sqrt{\epsilon_{p_{1}}}}} \left[ c_{1}^{\dagger} - c_{1} \right]
\end{equation}

\begin{equation}
q_{2}=\sqrt{\frac{\sqrt{\epsilon_{p_{2}}}}{2 \epsilon_{y}}} \left[ c_{2}^{\dagger} + c_{2} \right], \ p_{2}= i \sqrt{\frac{\epsilon_{y}}{2 \sqrt{\epsilon_{p_{2}}}}} \left[ c_{2}^{\dagger} - c_{2} \right]
\end{equation}

\begin{equation}
q_{3}=\sqrt{\frac{1}{2 \epsilon_{z}}} \left[ c_{3}^{\dagger} + c_{3} \right], \ p_{3}= i \sqrt{\frac{\epsilon_{z}}{2}} \left[ c_{3}^{\dagger} - c_{3} \right]
\end{equation}

We arrive at the final diagonal form as,

\begin{equation}\label{final_eqn}
H=\epsilon_{X} c_{1}^{\dagger} c_{1}+\epsilon_{Y}  c_{2}^{\dagger} c_{2} + \epsilon_{Z} c_{3}^{\dagger}c_{3} +\frac{1}{2} \left[  \epsilon_{X} + \epsilon_{Y} + \epsilon_{Z} \right]
\end{equation}

where, $\epsilon_{X}= \epsilon_{x} \sqrt{\epsilon_{p_{1}}}$, $\epsilon_{Y}= \epsilon_{y} \sqrt{\epsilon_{p_{2}}}$ and $\epsilon_{Z}= \epsilon_{z}$.

\begin{figure}[h]
\hspace{-0.0cm}
\begin{tabular}{c}
\includegraphics [scale=0.60]{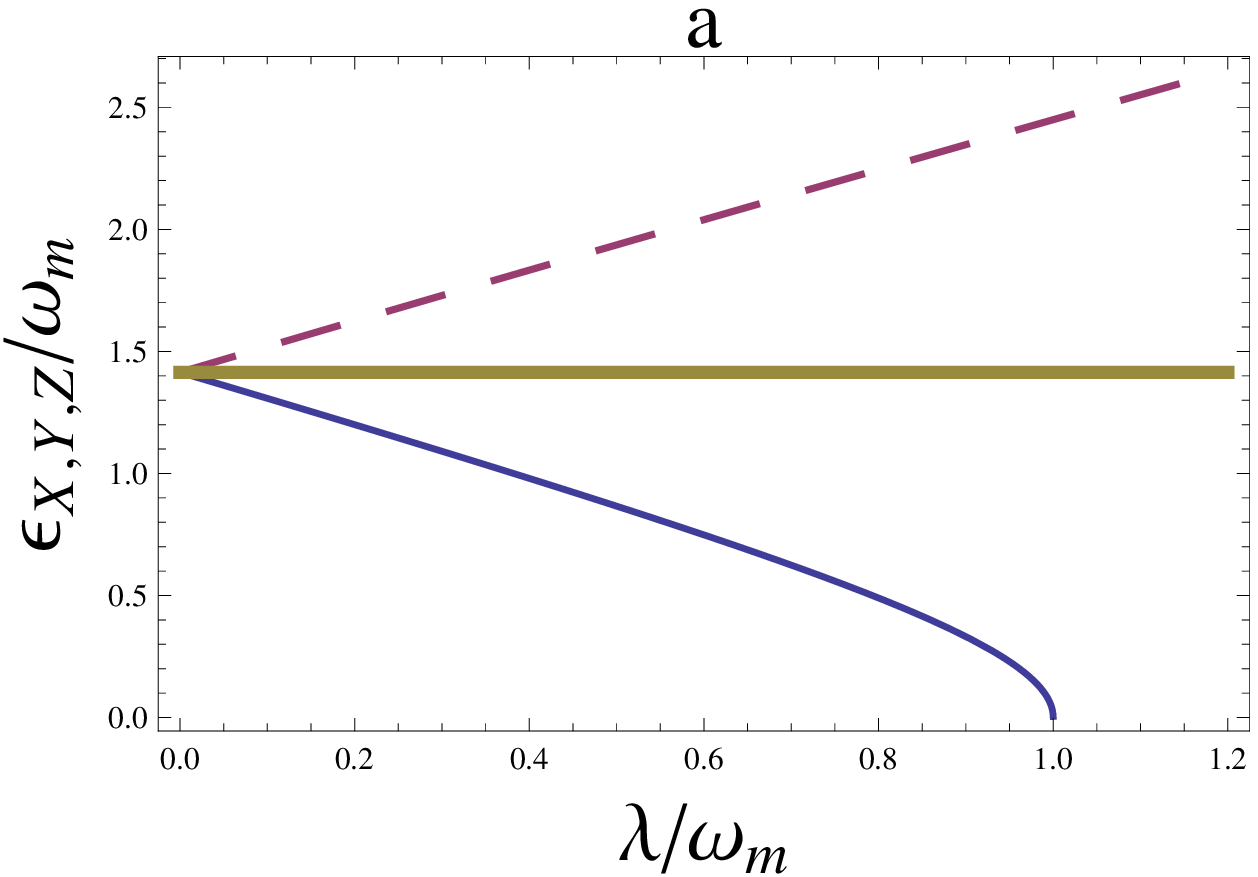}\\
\includegraphics [scale=0.60]{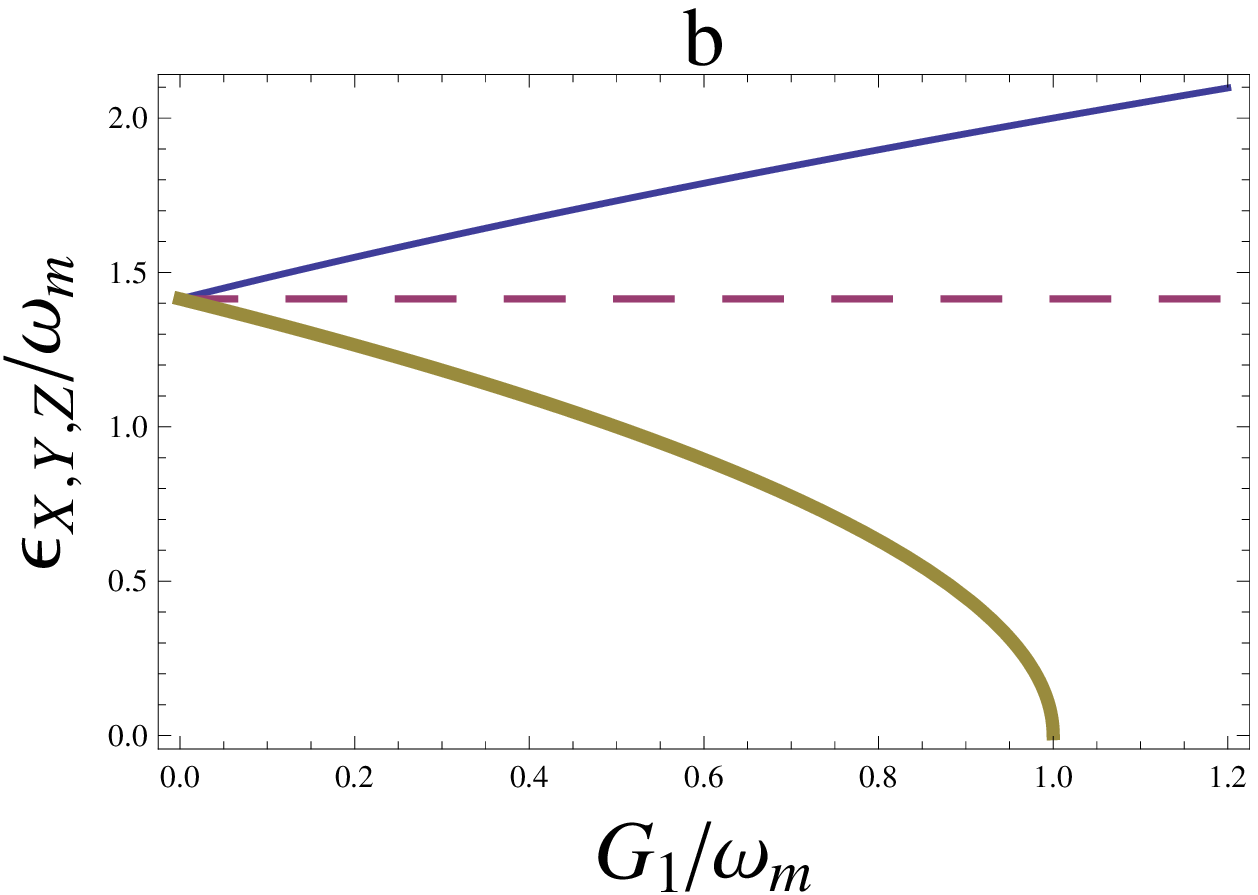}\\
\includegraphics [scale=0.60]{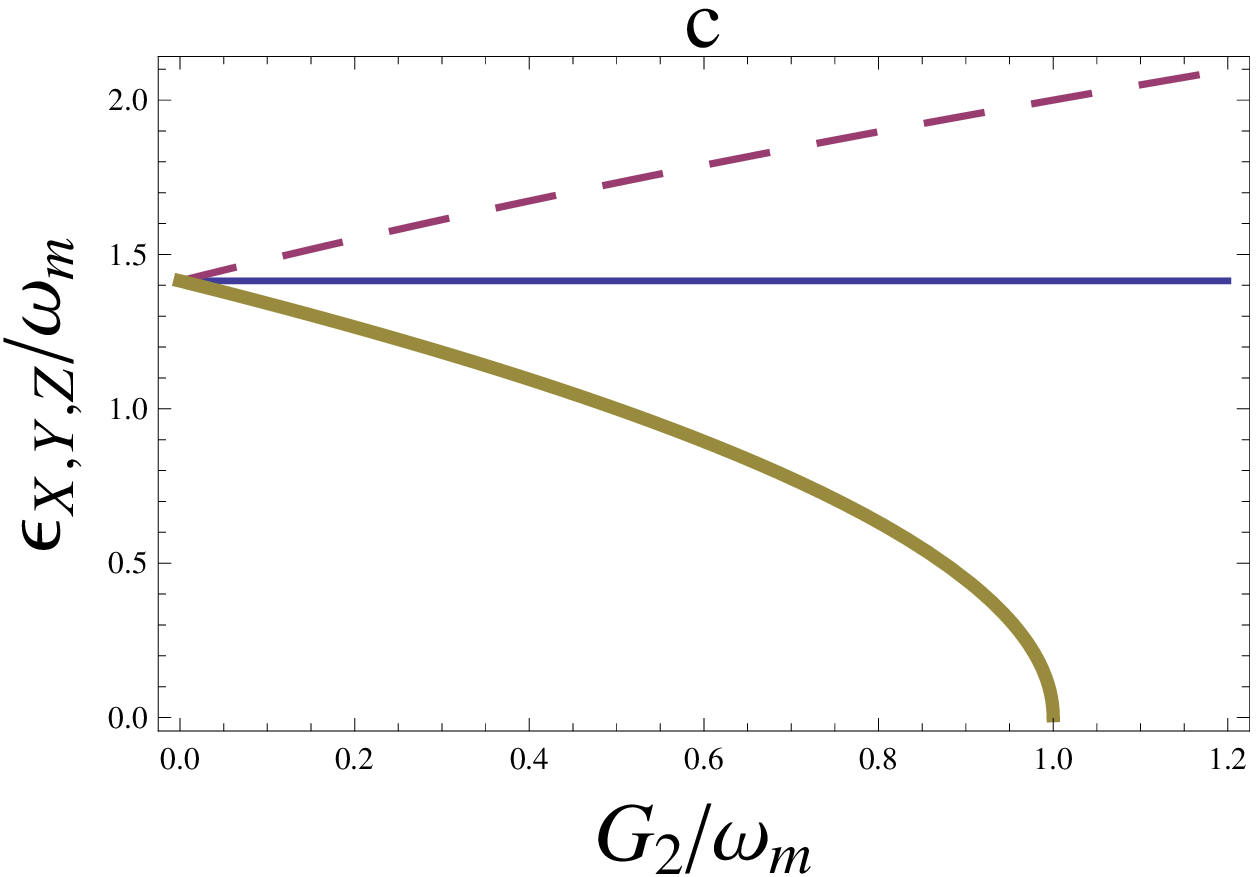}\\
\end{tabular}
\caption{(Color Online) The excitation energies $\epsilon_{X}$ (thin solid line), $\epsilon_{Y}$ (dashed line) and  $\epsilon_{Z}$ (thick solid line) of the 3-mode Dicke type Hamiltonian of equation \ref{final_eqn} as a function of coupling parameter $\lambda$ (plot (a)), $G_{1}$ (plot (b)) and $G_{2}$ (plot (c)). The $\epsilon_{X,Y}$ are the optical branches while $\epsilon_{Z}$ is the phononic branch. Selective energy transfer is seen in these plots. In plot (a), energy transfer between the $c_{1}$ optical mode and the $c_{2}$ optical mode is observed while the energy of the $c_{3}$ phononic mode does not change. For plot (b) energy exchange between the $c_{1}$ and $c_{3}$ modes is seen while in plot (c), energy exchange between $c_{2}$ and $c_{3}$ modes is clearly visible.}
\label{f1}
\end{figure}

It can be noted from equations \ref{eqn.x}, \ref{eqn.y}, \ref{eqn.z} and \ref{final_eqn}  that the value of the eigenvalue $\epsilon_{X}$ corresponding to the $c_{1}^{\dagger} c_{1}$ bosonic mode can become imaginary when $\epsilon_{x}^{2}\epsilon_{p_{1}} <0$. This happens when $\lambda > \sqrt{\Omega_{1} \Omega_{2}}$. This means that the system undergoes different bahaviours depending on being in the "normal phase" i.e when $\epsilon_{x}^{2} \epsilon_{p_{1}} > 0$ or in the so called "super-radiant phase" when  $\epsilon_{x}^{2} \epsilon_{p_{1}} < 0$. The eigenvalue $\epsilon_{Y}$ corresponding to the $c_{2}^{\dagger} c_{2}$ bosonic mode remains always real. The eigenvalue $\epsilon_{Z}$ can also become imaginary when we scan the coupling constants $G_{1}$ or $G_{2}$. Since we are interested only in the normal phase regime, we will not go into the details of the phase transition regime. The model described above is valid only in the normal phase. The excitation energies $\epsilon_{X}$ (thin solid line), $\epsilon_{Y}$ (dashed line) and  $\epsilon_{Z}$ (thick solid line) of the 3-mode Dicke type Hamiltonian of equation \ref{final_eqn} as a function of coupling parameter $\lambda$ (plot (a)), $G_{1}$ (plot (b)) and $G_{2}$ (plot (c)) is shown in Fig.1. The $\epsilon_{X,Y}$ are the optical branches while $\epsilon_{Z}$ is the phononic branch. Selective energy transfer is seen in these plots. In plot (a), energy transfer between the $c_{1}$ optical mode and the $c_{2}$ optical mode is observed while the energy of the $c_{3}$ phononic mode does not change. The eigenvalue $\epsilon_{X}$ displays the phase transition at the critical coupling constant $\lambda_{c}$. The critical coupling constant is $\lambda_{c}= \sqrt{\Omega_{1} \Omega_{2}}$. The critical point at $\lambda_{c}$ is a consequence of the coupling between the momentum quadratures of the two optical modes. This is a new feature that we observed in this system. Interestingly we found that for some parameters the eigenvalue $\epsilon_{X}$ takes negative values, signifying instability. This happens when both $\epsilon_{p_{1}} <0 $ and $\epsilon_{x}^{2} <0$.  The value of $\lambda_{us}$ above which the system is unstable is:

\begin{eqnarray}
\lambda_{us}=\frac{\sqrt{(2\Omega_{1}^{2}+\Omega_{2}^{2}+\omega_{m}^{2}+
\sqrt{(\Omega_{1}^{2}-\omega_{m}^{2})^{2}+16G_{1}^{2}\Omega_{1}\omega_{m}})^{2}
-(\Omega_{2}^{2}-\Omega_{1}^{2})^{2}}}{2\sqrt{\Omega_{1}\Omega_{2}}}.
\end{eqnarray}

 In absence of any momentum coupling, $\lambda_{us}$ is the usual critical point $\lambda_{c}$. For plot (b) energy exchange between the $c_{1}$ and $c_{3}$ modes is seen. Now in this case $\epsilon_{Z}$ demonstrates the phase transition at a certain critical coupling parameter $G_{1c}$. The value of $G_{1}$ above which the excitation energy $\epsilon_{Z}$ becomes imaginary is:

\begin{eqnarray}
G_{1}>\frac{\sqrt{(\Omega_{1}^{2}+\Omega_{2}^{2}+2\omega_{m}^{2}-
\sqrt{(\Omega_{2}^{2}-\omega_{m}^{2})^{2}+16G_{2}^{2}\Omega_{2}\omega_{m}})^{2}
-(\Omega_{1}^{2}-\omega_{m}^{2})^{2}}}{4\sqrt{\Omega_{1}\omega_{m}}}.
\end{eqnarray}

The superradiant phase transition in the phonon excitation spectrum indicates that complete energy transfer from the mechanical mode to the optical mode occurs at the quantum critical point. Similarly in plot (c), energy exchange between $c_{2}$ and $c_{3}$ modes is clearly visible.

\section{Mode Squeezing}

In this section, we investigate the squeezing behaviour of the three bosonic modes. A bosonic mode is said to be squeezed if the uncertainty in either of its position or momentum quadrature is less than the uncertainty in a coherent state \citep{walls}. A coherent state is a minimum uncertainty state which satisfies $(\Delta \alpha)^{2} (\Delta p_{\alpha})^{2} = 1/4 $ $[\alpha=x, y, z]$ and with uncertainties equally distributed between the two quadratures. A bosonic field is squeezed if $(\Delta \alpha)^{2} < 1/2 $ or $(\Delta p_{\alpha})^{2} < 1/2$ \citep{walls}. The two quadratures variances of the original field modes $a_{1}$, $a_{2}$ and $b$ are defined as $(\Delta \alpha)^{2} = <\alpha^{2}>- <\alpha>^{2}$ and $(\Delta p_{\alpha})^{2}= <p_{\alpha}^{2}>-<p_{\alpha}>^{2}$. The relationship between the original bosonic modes $[a_{1}, a_{2}, b]$ and the transformed bosonic modes $[c_{1}, c_{2}, c_{3}]$ are given as,

\begin{eqnarray}
a_{1}^{\dagger} &=& \frac{1}{2}  \{ \frac{\cos{\gamma_{1}}+\cos{\gamma_{2}}}{\sqrt{\Omega_{1} \epsilon_{x} \sqrt{\epsilon_{p_{1}}}}} \left[ c_{1}^{\dagger}(\sqrt{\epsilon_{p_{1}}} \Omega_{1}+\epsilon_{x})+ c_{1} (\sqrt{\epsilon_{p_{1}}} \Omega_{1}-\epsilon_{x}) \right] \nonumber \\
&+& \frac{\sin{\gamma_{1}}}{\sqrt{\sqrt{\epsilon_{p_{2}}} \Omega_{1} \epsilon_{y}}}  \left[ c_{2}^{\dagger}(\sqrt{\epsilon_{p_{2}}} \Omega_{1}+\epsilon_{y})+ c_{2} (\sqrt{\epsilon_{p_{2}}} \Omega_{1}-\epsilon_{y})\right ] \nonumber \\
&+& \frac{\sin{\gamma_{2}}}{\sqrt{\Omega_{1} \epsilon_{z}}} \left[ c_{3}^{\dagger} (\Omega_{1}+ \epsilon_{z})+ c_{3} (\Omega_{1}-\epsilon_{z})\right]              \}
\end{eqnarray}

\begin{eqnarray}
a_{2}^{\dagger} &=& \frac{1}{2}  \{ \frac{- \sin{\gamma_{1}}}{\sqrt{\Omega_{2} \epsilon_{x} \sqrt{\epsilon_{p_{1}}}}} \left[ c_{1}^{\dagger}(\sqrt{\epsilon_{p_{1}}} \Omega_{2}+\epsilon_{x})+ c_{1} (\sqrt{\epsilon_{p_{1}}} \Omega_{2}-\epsilon_{x}) \right] \nonumber \\
&+& \frac{\cos{\gamma_{1}}+\cos{\gamma_{3}}}{\sqrt{\sqrt{\epsilon_{p_{2}}} \Omega_{2} \epsilon_{y}}}  \left[ c_{2}^{\dagger}(\sqrt{\epsilon_{p_{2}}} \Omega_{2}+\epsilon_{y})+ c_{2} (\sqrt{\epsilon_{p_{2}}} \Omega_{2}-\epsilon_{y})\right ] \nonumber \\
&+& \frac{\sin{\gamma_{3}}}{\sqrt{\Omega_{2} \epsilon_{z}}} \left[ c_{3}^{\dagger} (\Omega_{2}+ \epsilon_{z})+ c_{3} (\Omega_{2}-\epsilon_{z})\right]              \}
\end{eqnarray}

\begin{eqnarray}
b^{\dagger} &=& \frac{1}{2}  \{ \frac{- \sin{\gamma_{2}}}{\sqrt{\omega_{m} \epsilon_{x} \sqrt{\epsilon_{p_{1}}}}} \left[ c_{1}^{\dagger}(\sqrt{\epsilon_{p_{1}}} \omega_{m}+\epsilon_{x})+ c_{1} (\sqrt{\epsilon_{p_{1}}} \omega_{m}-\epsilon_{x}) \right] \nonumber \\
&-& \frac {\sin{\gamma_{3}}}{\sqrt{\sqrt{\epsilon_{p_{2}}} \omega_{m} \epsilon_{y}}}  \left[ c_{2}^{\dagger}(\sqrt{\epsilon_{p_{2}}} \omega_{m}+\epsilon_{y})+ c_{2} (\sqrt{\epsilon_{p_{2}}} \omega_{m}-\epsilon_{y})\right ] \nonumber \\
&+& \frac{\cos{\gamma_{2}}+\cos{\gamma_{3}}}{\sqrt{\omega_{m} \epsilon_{z}}} \left[ c_{3}^{\dagger} (\omega_{m}+ \epsilon_{z})+ c_{3} (\omega_{m}-\epsilon_{z})\right]              \}
\end{eqnarray}

\begin{figure}[h]
\hspace{-0.0cm}
\begin{tabular}{c}
\includegraphics [scale=0.60]{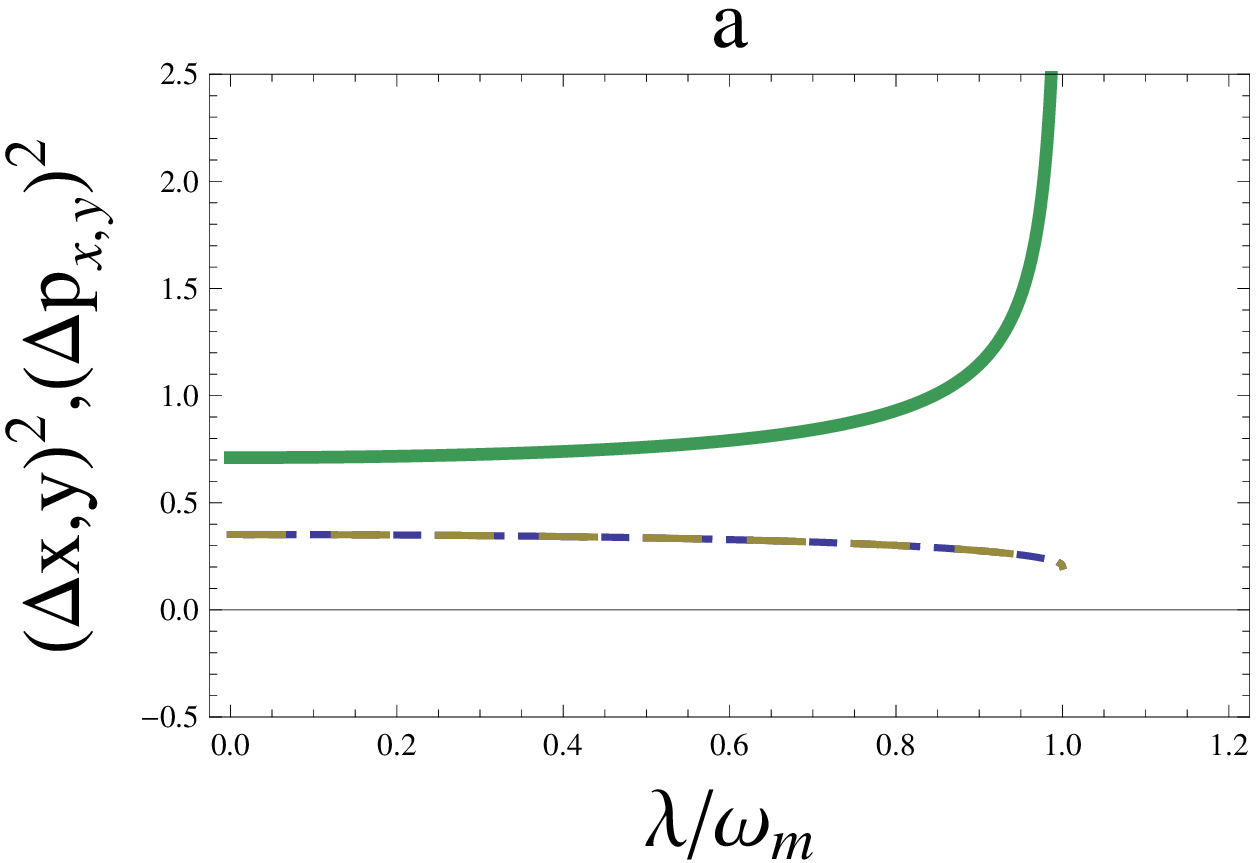}\\
\includegraphics [scale=0.60]{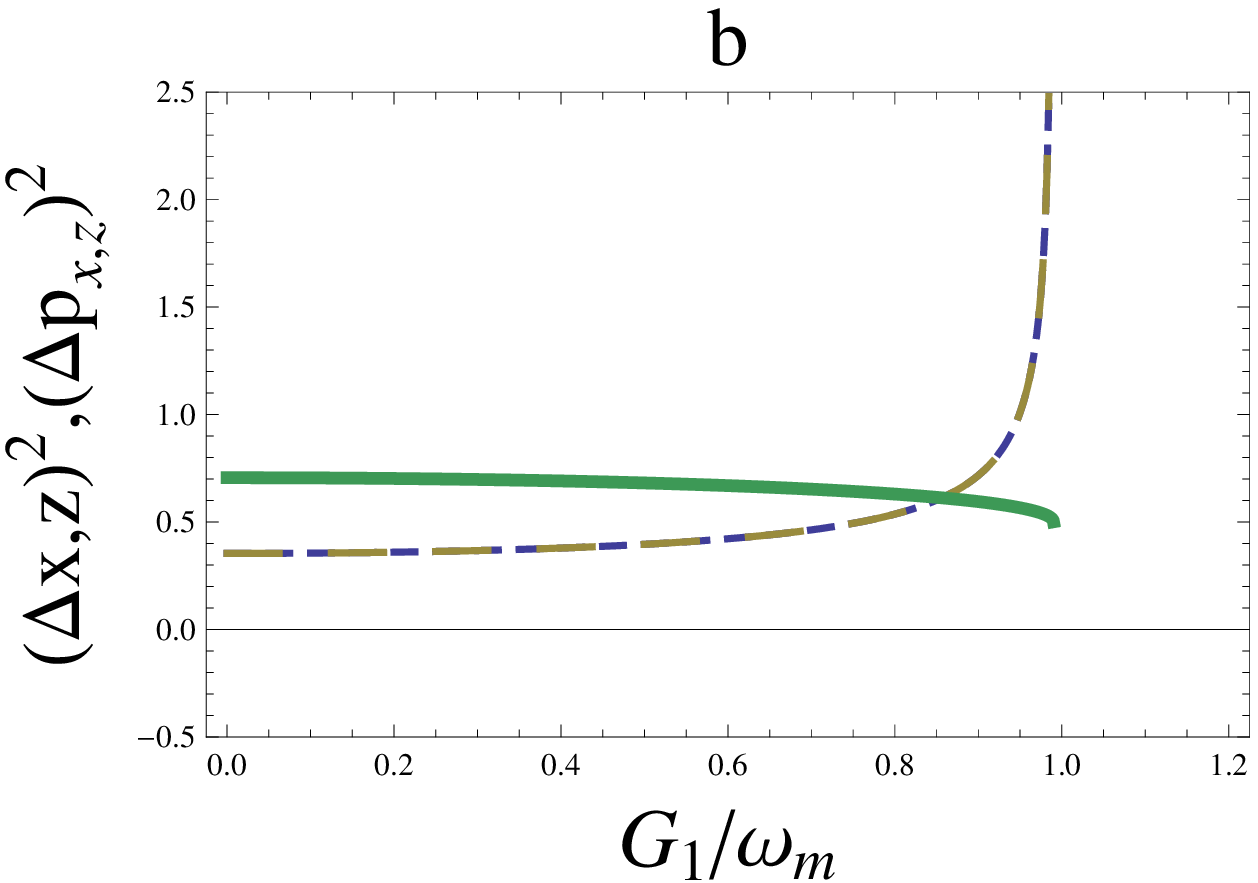}\\
\includegraphics [scale=0.60]{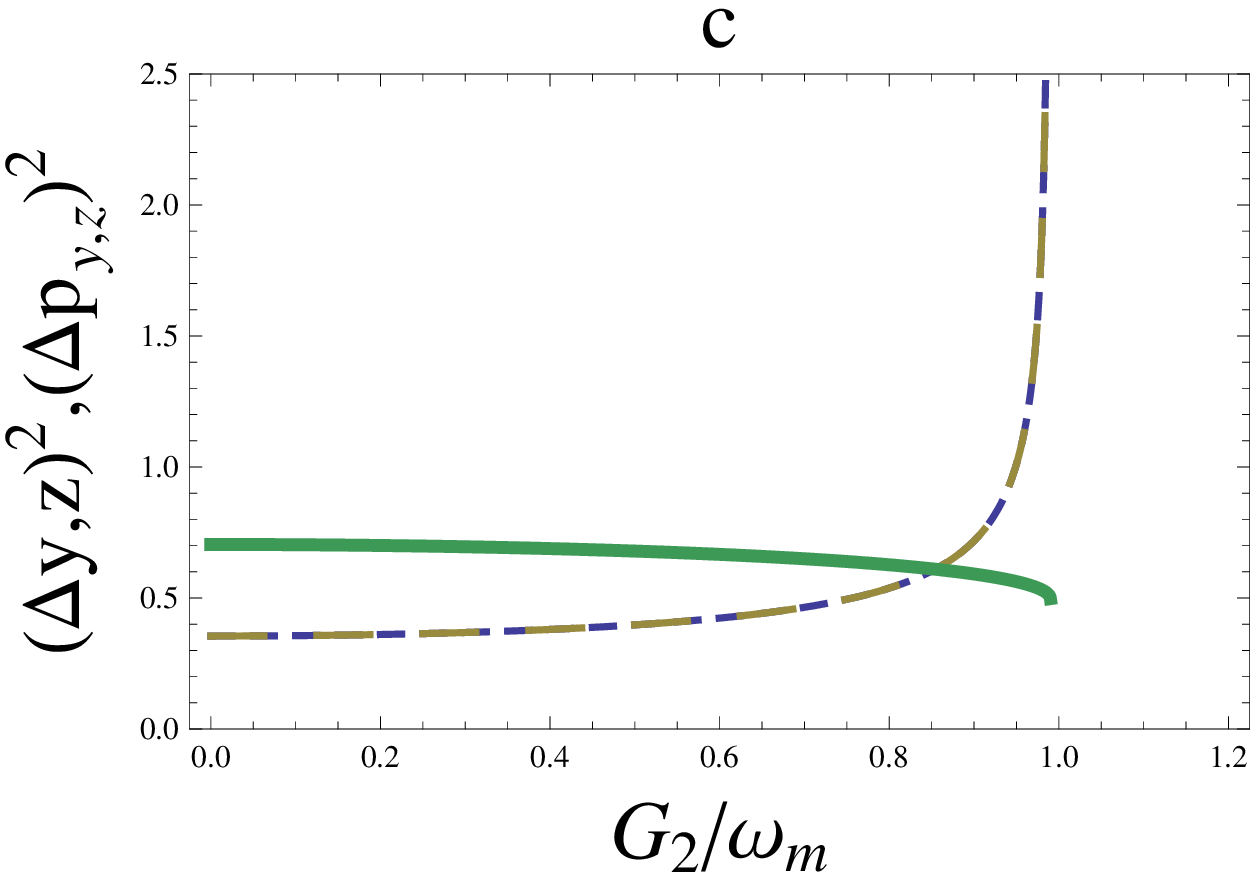}\\
\end{tabular}
\caption{The squeezing variances corresponding to Fig.2. (a): Plot of $(\Delta x)^{2}$ (dashed line)  ,$(\Delta y)^{2}$ (dashed line), $(\Delta p_{x})^{2}$ (solid line) and $(\Delta p_{y})^{2}$ (solid line) as a function of $\lambda/\omega_{m}$ for $\Omega_{1}=\omega_{m}$, $\Omega_{2}=\omega_{m}$, $G_{1}=0.01 \omega_{m}$ and $\Omega_{2}=0.01 \omega_{m}$. Note that $(\Delta x)^{2}$ and $(\Delta y)^{2}$ are coincident and the same for $(\Delta p_{x})^{2}$ and $(\Delta p_{y})^{2}$. (b): Plot of $(\Delta x)^{2}$ (dashed line)  ,$(\Delta z)^{2}$ (dashed line), $(\Delta p_{x})^{2}$ (solid line) and $(\Delta p_{z})^{2}$ (solid line) as a function of $G_{1}/\omega_{m}$ for $\Omega_{1}=\omega_{m}$, $\Omega_{2}=\omega_{m}$, $\lambda=0.01 \omega_{m}$ and $G_{2}=0.01 \omega_{m}$. (c): Plot of $(\Delta y)^{2}$ (dashed line)  ,$(\Delta z)^{2}$ (dashed line), $(\Delta p_{y})^{2}$ (solid line) and $(\Delta p_{z})^{2}$ (solid line) as a function of $G_{2}/\omega_{m}$ for $\Omega_{1}=\omega_{m}$, $\Omega_{2}=\omega_{m}$, $\lambda=0.01 \omega_{m}$ and $G_{1}=0.01 \omega_{m}$}.
\label{f2}
\end{figure}

One can trivially show that the various variances are,

\begin{equation}
(\Delta x)^{2}= \frac{1}{2 \Omega_{1}} \left\{ 1+ \frac{(\cos{\gamma_{1}}+\cos{\gamma_{2}})^{2}}{\epsilon_{x}} (\sqrt{\epsilon_{p_{1}}} \Omega_{1}- \epsilon_{x}) + \frac{\sin^{2}{\gamma_{1}}}{\epsilon_{y}} (\sqrt{\epsilon_{p_{2}}} \Omega_{1}- \epsilon_{y})+ \frac{\sin^{2}{\gamma_{2}}}{\epsilon_{z}} (\Omega_{1}-\epsilon_{z})      \right\}
\end{equation}

\begin{equation}
(\Delta y)^{2}= \frac{1}{2 \Omega_{2}} \left\{ 1+ \frac{\sin^{2}{\gamma_{1}}}{\epsilon_{x}} (\sqrt{\epsilon_{p_{1}}} \Omega_{2}- \epsilon_{x}) + \frac{(\cos{\gamma_{1}}+\cos{\gamma_{3}})^{2}}{\epsilon_{y}} (\sqrt{\epsilon_{p_{2}}} \Omega_{2}- \epsilon_{y})+ \frac{\sin^{2}{\gamma_{3}}}{\epsilon_{z}} (\Omega_{2}-\epsilon_{z})      \right\}
\end{equation}

\begin{equation}
(\Delta z)^{2}= \frac{1}{2 \omega_{m}} \left\{ 1+ \frac{\sin^{2}{\gamma_{2}}}{\epsilon_{x}} (\sqrt{\epsilon_{p_{1}}} \omega_{m}- \epsilon_{x}) + \frac{\sin^{2}{\gamma_{3}}}{\epsilon_{y}} (\sqrt{\epsilon_{p_{2}}} \omega_{m}- \epsilon_{y})+ \frac{(\cos{\gamma_{2}}+\cos{\gamma_{3}})^{2} }{\epsilon_{z}} (\omega_{m}-\epsilon_{z})      \right\}
\end{equation}

\begin{equation}
(\Delta p_{x})^{2}= \frac{\Omega_{1}}{2} \left\{ 1+ \frac{(\cos{\gamma_{1}}+\cos{\gamma_{2}})^{2}}{\sqrt{\epsilon_{p_{1}}} \Omega_{1}} (\epsilon_{x}-\sqrt{\epsilon_{p_{1}}} \Omega_{1}) + \frac{\sin^{2}{\gamma_{1}}}{\sqrt{\epsilon_{p_{2}}} \Omega_{1}} (\epsilon_{y}-\sqrt{\epsilon_{p_{2}}} \Omega_{1} )+ \frac{\sin^{2}{\gamma_{2}}}{\Omega_{1}} (\epsilon_{z}-\Omega_{1})      \right\}
\end{equation}

\begin{equation}
(\Delta p_{y})^{2}= \frac{\Omega_{2}}{2} \left\{ 1+ \frac{\sin^{2}{\gamma_{1}}}{\sqrt{\epsilon_{p_{1}}} \Omega_{2}} (\epsilon_{x}-\sqrt{\epsilon_{p_{1}}} \Omega_{2}) + \frac{(\cos{\gamma_{1}}+\cos{\gamma_{3}})^{2}}{\sqrt{\epsilon_{p_{2}}} \Omega_{2}} (\epsilon_{y}-\sqrt{\epsilon_{p_{2}}} \Omega_{2} )+ \frac{\sin^{2}{\gamma_{3}}}{\Omega_{2}} (\epsilon_{z}-\Omega_{2})      \right\}
\end{equation}

\begin{equation}
(\Delta p_{z})^{2}= \frac{\omega_{m}}{2} \left\{ 1+ \frac{\sin^{2}{\gamma_{2}}}{\sqrt{\epsilon_{p_{1}}} \omega_{m}} (\epsilon_{x}-\sqrt{\epsilon_{p_{1}}} \omega_{m}) + \frac{\sin^{2}{\gamma_{3}}}{\sqrt{\epsilon_{p_{2}}} \omega_{m}} (\epsilon_{y}-\sqrt{\epsilon_{p_{2}}} \omega_{m} )+ \frac{(\cos{\gamma_{2}}+\cos{\gamma_{3}})^{2}}{\omega_{m}} (\epsilon_{z}-\omega_{m})      \right\}
\end{equation}

Fig.3 displays the squeezing variances corresponding to Fig.2. In Fig.3(a), the momentum variances $(\Delta p_{x})^{2}$ and $(\Delta p_{y})^{2}$ are unsqueezed and diverges (more unsqueezed) as $\lambda$ approaches $\lambda_{c}$. On the other hand the position variances $(\Delta x)^{2}$ and $(\Delta y)^{2}$ are already squeezed and become more squeezed as $\lambda$ approaches $\lambda_{c}$. In contrast, the Figures 3(b) and 3(c) illustrates that the momentum variances which are initially unsqueezed, appoaches the value $1/2$ and becomes slightly squeezed as the coupling parameters approach the critical value. The corresponding position variances which are initially squeezed diverges and becomes unsqueezed as the critical point is approached. Thus we conclude that the quadratures which are squeezed are best suited for quantum measurements since squeezed quadratures have reduced quantum noise i.e, if we make quantum measurements while varying $\lambda$, then it would be more appropriate to make measurements of the position quadratures. On the other hand, while varying $G_{1}$ or $G_{2}$,the position quadrature is a better candidate as long as it squeezed. Near the critical point, measurements on the momentum quadrature would be better suited.

\section{Displacement Spectrum: Normal Mode Splitting}

Here we show that coupling between the mechanical mode fluctuations and the two cavity mode fluctuations leads to a splitting of the normal mode into three modes (Normal mode splitting (NMS)).The optomechanical NMS however involves driving three parametrically coupled nondegenerate modes out of equilibrium.
The NMS does not appear in the steady state spectra but rather manifests itself in the fluctuation spectra of the mirror displacement.

Using the Hamiltonian given by Eqn. (\ref{ham2}), we get the linearized equations of motion which are represented as follows:

\begin{equation}
\dot{a_{1}}(t)= [i\Delta_{1}-\frac{\gamma_{c_{1}}}{2}]a_{1}(t)+iG_{1}(b(t)+b^{\dagger}(t))- i\lambda a_{2}(t)+\sqrt{\gamma_{c_{1}}} a_{in_{1}}(t),
\end{equation}

\begin{equation}
\dot{a_{2}}(t)= [i\Delta_{2}-\frac{\gamma_{c_{2}}}{2}]a_{2}(t)+iG_{2}(b(t)+b^{\dagger}(t))- i\lambda a_{1}(t)+\sqrt{\gamma_{c_{2}}} a_{in_{2}}(t),
\end{equation}

\begin{equation}
\dot{b}(t)= -(i\omega_{m}+\gamma_{m})b(t)+iG_{1}(a_{1}(t)+a_{1}^{\dagger}(t))+iG_{2}(a_{2}(t)+a_{2}^{\dagger}(t))+\sqrt{\gamma_{m}}\xi(t),
\end{equation}

where $\Delta_{1}=-\Omega_{1}$, $G_{1}=g_{1} \alpha_{1}-G \alpha_{2}$, $\Delta_{2}=-\Omega_{2}$ and $G_{2}=g_{2} \alpha_{2}-G \alpha_{1}$. Now, we rewrite the above equations in terms of amplitude and phase quadratures for the system with $X_{1}(t)=[a_{1}(t)+a_{1}^{\dagger}(t)]$, $Y_{1}(t)=i[a_{1}^{\dagger}(t)-a_{1}(t)]$, $X_{2}(t)=[a_{2}(t)+a_{2}^{\dagger}(t)]$, $Y_{2}(t)=i[a_{2}^{\dagger}(t)-a_{2}(t)]$, $Q(t)=[b(t)+b^{\dagger}(t)]$, $P(t)=i[b^{\dagger}(t)-b(t)]$, $X_{in_{1}}(t)=[a_{in_{1}}(t)+a_{in_{1}}^{\dagger}(t)]$, $Y_{in_{1}}(t)=i[a_{in_{1}}^{\dagger}(t)-a_{in_{1}}(t)]$, $X_{in_{2}}(t)=[a_{in_{2}}(t)+a_{in_{2}}^{\dagger}(t)]$ and $Y_{in_{2}}(t)=i[a_{in_{2}}^{\dagger}(t)-a_{in_{2}}(t)]$ as:

\begin{equation}
\dot{X_{1}}(t)=-\Delta_{1}Y_{1}(t)+\lambda Y_{2}(t)-\frac{\gamma_{c_{1}}}{2}X_{1}(t)+\sqrt{\gamma_{c_{1}}} X_{in_{1}}(t),
\end{equation}

\begin{equation}
\dot{Y_{1}}(t)=\Delta_{1} X_{1}(t)+G_{1} Q(t)-\lambda X_{2}(t)+\sqrt{\gamma_{c_{1}}} Y_{in_{1}}(t)-\frac{\gamma_{c_{1}}}{2} Y_{1}(t),
\end{equation}

\begin{equation}
\dot{X_{2}}(t)=-\Delta_{2} Y_{2}(t)+\lambda Y_{1}(t)-\frac{\gamma_{c_{2}}}{2} X_{2}(t)+\sqrt{\gamma_{c_{2}}}X_{in_{2}}(t),
\end{equation}

\begin{equation}
\dot{Y_{2}}(t)=\Delta_{2} X_{2}(t)+G_{2} Q(t)-\lambda X_{1}(t)-\frac{\gamma_{c_{2}}}{2} Y_{2}(t)+\sqrt{\gamma_{c_{2}}}Y_{in_{2}}(t),
\end{equation}

\begin{equation}
\dot{Q}(t)=\omega_{m} P(t),
\end{equation}

\begin{equation}
\dot{P}(t)=-\omega_{m} Q(t)-\gamma_{m} P(t)+G_{1} X_{1}(t)+G_{2} X_{2}(t)+W(t).
\end{equation}

Here $W(t)=i\sqrt{\gamma_{m}}(\xi^{\dagger}(t)-\xi(t))$ satisfies the following correlation \citep{gio}:

\begin{equation}
\left\langle W(t)W(t') \right\rangle = \frac{\gamma_{m}}{\omega_{m}} \int \frac{d\omega}{2\pi} e^{-i\omega(t-t')} \omega\left[1+\coth\left(\frac{\hbar\omega}{2k_{B}T} \right)  \right],
\end{equation}

where $k_{B}$ is the Boltzmann constant and $T$ is the temperature of the reservoir.

\begin{figure}[h]
\hspace{-0.0cm}
\includegraphics [scale=1]{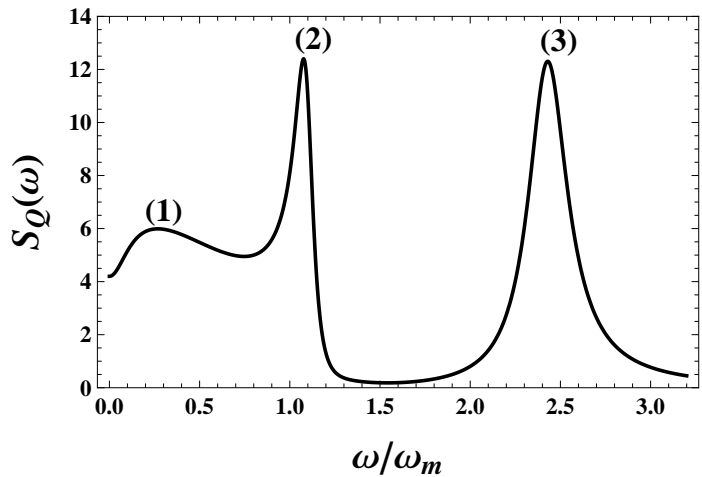}
\caption{Plot of displacement spectrum ($S_{Q}(\omega)$) as a function of dimensionless frequency($\omega/\omega_{m}$).Parameter values are : $\gamma_{m}=10^{-4}\omega_{m}$, $\Delta_{1}=-1.3\omega_{m}$, $\Delta_{2}=-1.5\omega_{m}$, $\gamma_{c_{1}}=0.2\omega_{m}$, $\gamma_{c_{2}}=0.6\omega_{m}$, $G=1.5\omega_{m}$, $\beta=0.06$, $G_{1}=2\omega_{m}$, $G_{2}=6\omega_{m}$ and $k_{B}T/\hbar\omega_{m}=10^{5}$.}
\label{1}
\end{figure}

The displacement spectrum for the mechanical mode of the system is obtained by using

\begin{equation}
S_{Q}(\omega)=\frac{1}{4\pi} \int d\omega'  e^{-i(\omega+\omega')t} \left\langle  Q(\omega) Q(\omega')+ Q(\omega') Q{\omega}\right\rangle.
\end{equation}

where
\begin{equation}\label{Q}
 Q(\omega)=\frac{A_{1}(\omega)+A_{2}(\omega)+A_{3}(\omega)+A_{4}(\omega)+A_{5}(\omega)}{B(\omega)}
\end{equation}

using the correlation functions given in the Appendix II. The values of $A_{1}(\omega)$, $A_{2}(\omega)$, $A_{3}(\omega)$, $A_{4}(\omega)$, $A_{5}(\omega)$ and $B(\omega)$ are given in the Appendix III.

Fig.3 shows the displacement spectrum ($S_{Q}(\omega)$) as a function of dimensionless frequency($\omega/\omega_{m}$). The NMS is associated with the mixing between the mechanical mode and the fluctuation of the two cavity field around the steady state. Clearly we see three modes in the displacement spectrum indicating a coherent energy exchange between the mechanical mode and the two optical modes. An important point to note is that in order to
observe the NMS, the energy exchange between the three modes should take place on a time scale faster than the decoherence of each mode. On the positive
detuning side, the observation of NMS is prevented by the onset of parametric instability.

The experimental prospects for various parameters used in the main paper are illustrated as follows. Mechanical frequency of optomechanical system can take the value $2 \pi \times 73.5$ MHz with corresponding damping rate $2 \pi \times 1.3$ KHz \citep{sch}. High Finesse optical cavities can decay with a rate nearly $2 \pi \times 10$ MHz \citep{tanaka,notomi}. Optomechanical crystal setups record a coupling rate of $2 \pi \times 1$ MHz \citep{chan1}. Such coupling rates can be enhanced above $10$ MHz by using nanoslots \citep{robin} which increases the local cavity field in these structures. The limit $\hbar\omega_{m}<<\hbar\gamma_{m}<<k_{B}T$ is always taken into account for various optomechanical experiments \citep{hadjar,titt,cohadon,pinard}.

\section{conclusion}

We have presented how the existence of a quantum phase transition in a hybrid optomechanical system can be exploited to selectively transfer energy between two optical modes and a single mechanical mode. By scanning one of the three coupling parameters, we can selectively open one channel to transfer energy between any two modes while at the same time close the other two channels. At the quantum critical point a complete transfer of energy is possible between any two modes. The study on the momentum and position variances tells us that the quadratures which are squeezed are best suited for measurements due to less amount of quantum noise. The normal mode splitting shows distinct three peaks which again demonstrates that the process of coherent energy exchange between the three modes is possible. The fact that phononic modes can store energy for a longer duration and photonic modes can transfer energy over long distances makes this hybrid optomechanical system useful in next generation of quantum communications and quantum information processing units.

\section{Appendix I}

Starting from the Hamiltonian of Eqn.(\ref{ham1}), we derive the Heisenberg equations of motion for the operators $a_{1}(t)$, $a_{2}(t)$ and $b(t)$. This yields,

\begin{equation}\label{a1}
\dot{a_{1}}(t)= -i \omega_{1} a_{1}(t)+ i g_{1} a_{1}(t) (b(t)+b^{\dagger}(t))-i G a_{2}(t) (b(t)+b^{\dagger}(t)) -\frac{\gamma_{c_{1}}}{2}a_{1}(t)+ \sqrt{\gamma_{c_{1}}} a_{in_{1}}(t),
\end{equation}

\begin{equation}\label{a2}
\dot{a_{2}}(t)=-i\omega_{2}a_{2}(t)+ig_{2}a_{2}(t)(b(t)+b^{\dagger}(t))-iGa_{1}(t)(b(t)+b^{\dagger}(t))-\frac{\gamma_{c_{2}}}{2}a_{2}(t)+\sqrt{\gamma_{c_{2}}}a_{in_{2}}(t),
\end{equation}

\begin{equation}\label{b}
\dot{b}(t)=-i\omega_{m}b(t)+ig_{1}a_{1}^{\dagger}(t)a_{1}(t)+ig_{2}a_{2}^{\dagger}(t)a_{2}(t)-iG(a_{1}(t)a_{2}^{\dagger}(t)+a_{2}(t)a_{1}^{\dagger}(t))-\gamma_{m}b(t)+\sqrt{\gamma_{m}}\xi(t).
\end{equation}

where $a_{in_{1}}(t)$ and $a_{in_{2}}(t)$ are the input noise operators for the two optical modes. $\xi(t)$ is the noise operator arising from the brownian motion of the mechanical mode. Now, we find the steady state parameters for the different operators  by factorizing the non-linear algebraic eqs. (\ref{a1})-(\ref{b}) and setting their time derivatives to zero, given as:

\begin{equation}
\alpha_{1}=\frac{iG \alpha_{2}(\beta+\beta^{\dagger})}{\left[ ig_{1}(\beta+\beta^{\dagger})-(i\omega_{1}+\frac{\gamma_{c_{1}}}{2})\right] },
\end{equation}

\begin{equation}
\alpha_{2}=\frac{iG \alpha_{1}(\beta+\beta^{\dagger})}{\left[ ig_{2}(\beta+\beta^{\dagger})-(i\omega_{2}+\frac{\gamma_{c_{2}}}{2})\right] },
\end{equation}

\begin{equation}
\beta=\frac{ig_{1} \alpha_{1}^{\dagger} \alpha_{1}+ig_{2} \alpha_{2}^{\dagger} \alpha_{2}-iG( \alpha_{1} \alpha_{2}^{\dagger}+\alpha_{2} \alpha_{1}^{\dagger})}{(i\omega_{m}+\gamma_{m})}.
\end{equation}

where $\alpha_{1}$, $\alpha_{2}$ and $\beta$ are the steady state values for the two optical cavity modes and the mechanical mode respectively.

\section{Appendix II}

The correlation function for the noise operator arising from brownian motion in the fourier space is given by \citep{tom}:

\begin{equation}
\left\langle W(\omega)W(\omega')\right\rangle=2\pi \omega \frac{\gamma_{m}}{\omega_{m}}\left\lbrace 1+\coth\left(\frac{\hbar \omega}{2 k_{B}T} \right)\right\rbrace \delta(\omega+\omega').
\end{equation}

The correlation functions for the amplitude and phase quadratures of the various input noise operators in the fourier space are given as follows \citep{tom}:

\begin{equation}
\left\langle X_{in_{1}}(\omega) X_{in_{1}}(\omega')\right\rangle
=\left\langle X_{in_{2}}(\omega) X_{in_{2}}(\omega')\right\rangle
=\left\langle Y_{in_{1}}(\omega) Y_{in_{1}}(\omega')\right\rangle
=\left\langle Y_{in_{2}}(\omega) Y_{in_{2}}(\omega')\right\rangle=2 \pi \delta(\omega+\omega') ,
\end{equation}

\begin{equation}
\left\langle X_{in_{1}}(\omega) Y_{in_{1}}(\omega')\right\rangle =
\left\langle X_{in_{2}}(\omega) Y_{in_{2}}(\omega')\right\rangle = 2 i \pi \delta(\omega+\omega'),
\end{equation}

\begin{equation}
\left\langle Y_{in_{1}}(\omega) X_{in_{1}}(\omega')\right\rangle =
\left\langle Y_{in_{2}}(\omega) X_{in_{2}}(\omega')\right\rangle = -2i \pi \delta(\omega+\omega').
\end{equation}

\section{Appendix III}

The coefficients in equation (\ref{Q}) are given as follows:

\begin{equation}
A_{1}(\omega)=W(\omega)\omega_{m}\left(\Delta_{2}^{2}-\omega^{2}+\frac{\gamma_{c_{2}}^{2}}{4}+i\omega\gamma_{c_{2}} \right)C_{1}(\omega)C_{2}(\omega),
\end{equation}

\begin{equation}
A_{2}(\omega)=X_{in_{1}}(\omega)\sqrt{\gamma_{c_{1}}}\left(\Delta_{2}^{2}-\omega^{2}+\frac{\gamma_{c_{2}}^{2}}{4}+i\omega\gamma_{c_{2}} \right)C_{1}(\omega)C_{3}(\omega),
\end{equation}

\begin{eqnarray}\nonumber
A_{3}(\omega)=Y_{in_{1}}(\omega)\sqrt{\gamma_{c_{1}}}\left(\Delta_{2}^{2}-\omega^{2}+\frac{\gamma_{c_{2}}^{2}}{4}+i\omega\gamma_{c_{2}} \right)\left[\lambda\omega_{m}G_{2}\left(i\omega+\frac{\gamma_{c_{2}}}{2}\right) C_{2}(\omega)\right.\\\left. +
\left\lbrace \lambda^{2}\Delta_{2}-\Delta_{1}\left(\Delta_{2}^{2}-\omega^{2}+\frac{\gamma_{c_{2}}^{2}}{4}+i\omega\gamma_{c_{2}} \right) \right\rbrace C_{3}(\omega)\right] ,
\end{eqnarray}

\begin{eqnarray}\nonumber
A_{4}(\omega)=X_{in_{2}}(\omega)\sqrt{\gamma_{c_{2}}}\left(\Delta_{2}^{2}-\omega^{2}+\frac{\gamma_{c_{2}}^{2}}{4}+i\omega\gamma_{c_{2}} \right)\left[\lambda^{2}\omega_{m}G_{2}\left\lbrace \Delta_{1}\left(i\omega+\frac{\gamma_{c_{2}}}{2} \right)+\Delta_{2}\left(i\omega+\frac{\gamma_{c_{1}}}{2} \right) \right\rbrace^{2}\right.\\\left. +\omega_{m}G_{2}C_{2}(\omega)\left(i\omega +\frac{\gamma_{c_{1}}}{2}\right)\left(i\omega +\frac{\gamma_{c_{2}}}{2}\right)+\lambda\omega_{m}G_{1}C_{1}(\omega)\left\lbrace \Delta_{1}\left(i\omega+\frac{\gamma_{c_{2}}}{2} \right)+\Delta_{2}\left(i\omega+\frac{\gamma_{c_{1}}}{2} \right) \right\rbrace \right],
\end{eqnarray}

\begin{eqnarray}\nonumber
A_{5}(\omega)=Y_{in_{2}}(\omega)\sqrt{\gamma_{c_{2}}}\left[\lambda C_{3}(\omega)\left\lbrace \lambda^{2}\Delta_{2}^{2}-\Delta_{1}\Delta_{2}\left(\Delta_{2}^{2}-\omega^{2}+\frac{\gamma_{c_{2}}^{2}}{4}+i\omega\gamma_{c_{2}} \right)+C_{1}(\omega)\left(i\omega+\frac{\gamma_{c_{2}}}{2} \right) \right\rbrace\right.\\\left. -\omega_{m}G_{2}\Delta_{2}C_{2}(\omega)\left(i\omega+\frac{\gamma_{c_{1}}}{2} \right)\left(\Delta_{2}^{2}-\omega^{2}+\frac{\gamma_{c_{2}}^{2}}{4}+i\omega\gamma_{c_{2}} \right) \right],
\end{eqnarray}

\begin{equation}
B(\omega)=C_{4}(\omega)-C_{5}(\omega),
\end{equation}

where

\begin{equation}
C_{1}(\omega)=\left( i\omega+\frac{\gamma_{c_{1}}}{2}\right) \left(\Delta_{2}^{2}-\omega^{2}+\frac{\gamma_{c_{2}}^{2}}{4}+i\omega\gamma_{c_{2}} \right)+\lambda^{2}\left( i\omega+\frac{\gamma_{c_{2}}}{2}\right),
\end{equation}

\begin{eqnarray}
C_{2}(\omega)&=& \left(\Delta_{2}^{2}-\omega^{2}+\frac{\gamma_{c_{2}}^{2}}
{4}+i\omega\gamma_{c_{2}}
\right)\left(\Delta_{1}^{2}-\omega^{2}+\frac{\gamma_{c_{1}}^{2}}{4}+i\omega\gamma_{c_{1}} \right)+\lambda^{4} \\ \nonumber
&+& 2\lambda^{2}\left\lbrace \left( i\omega+\frac{\gamma_{c_{1}}}{2}\right) \left( i\omega+\frac{\gamma_{c_{2}}}{2}\right) -\Delta_{1}\Delta_{2}\right\rbrace,
\end{eqnarray}

\begin{eqnarray}
C_{3}(\omega)&=& \omega_{m}G_{1}\left\lbrace \left( i\omega+\frac{\gamma_{c_{1}}}{2}\right) \left(\Delta_{2}^{2}-\omega^{2}+\frac{\gamma_{c_{2}}^{2}}{4}+i\omega\gamma_{c_{2}} \right)+\lambda^{2}\left( i\omega+\frac{\gamma_{c_{2}}}{2}\right)\right\rbrace \\ \nonumber
&+& \lambda\omega_{m}G_{2}\Delta_{1}\left(i\omega+\frac{\gamma_{c_{2}}}{2} \right) +\lambda\omega_{m}G_{2}\Delta_{2}\left(i\omega+\frac{\gamma_{c_{1}}}{2} \right),
\end{eqnarray}

\begin{eqnarray}\nonumber
C_{4}(\omega)=\left(\Delta_{2}^{2}-\omega^{2}+\frac{\gamma_{c_{2}}^{2}}{4}+i\omega\gamma_{c_{2}} \right)C_{2}(\omega)\left[(\omega_{m}^{2}-\omega^{2}+i\omega\gamma_{m})C_{1}(\omega)+\omega_{m}G^{2}\Delta_{2}\left( i\omega+\frac{\gamma_{c_{1}}}{2}\right)\right.\\\left.-\lambda\omega_{m}G_{1}G_{2}\left( i\omega+\frac{\gamma_{c_{2}}}{2}\right)\right],
\end{eqnarray}

\begin{eqnarray}\nonumber
C_{5}(\omega)=C_{3}(\omega)\left[\lambda G_{2}C_{1}(\omega)\left( i\omega+\frac{\gamma_{c_{2}}}{2}\right)-\left\lbrace \lambda G_{2}\Delta_{2}+G_{1}\left(\Delta_{2}^{2}-\omega^{2}+\frac{\gamma_{c_{2}}^{2}}{4}+i\omega\gamma_{c_{2}} \right)\right\rbrace\right.\\\left.\left\lbrace \Delta_{1}\left(\Delta_{2}^{2}-\omega^{2}+\frac{\gamma_{c_{2}}^{2}}{4}+i\omega\gamma_{c_{2}} \right)-\lambda^{2} \Delta_{2}\right\rbrace\right].
\end{eqnarray}

\section{Acknowledgements}
 The authors acknowledge financial support from the Department of Science and Technology, New Delhi for financial assistance vide grant SR/S2/LOP-0034/2010.

\end{document}